# Taming the Yukawa potential singularity: improved evaluation of bound states and resonance energies

A. D. Alhaidari[*]
*Shura Council, Riyadh 11212, Saudi Arabia*

H. Bahlouli and M. S. Abdelmonem
*Physics Department, King Fahd University of Petroleum & Minerals, Dhahran 31261, Saudi Arabia*

Using the tools of the J-matrix method, we absorb the 1/r singularity of the Yukawa potential in the reference Hamiltonian, which is handled analytically. The remaining part, which is bound and regular everywhere, is treated by an efficient numerical scheme in a suitable basis using Gauss quadrature approximation. Analysis of resonance energies and bound states spectrum is performed using the complex scaling method, where we show their trajectories in the complex energy plane and demonstrate the remarkable fact that bound states cross over into resonance states by varying the potential parameters.

The Yukawa potential [1] is used in various areas of physics to model singular but short-range interactions. In high energy physics, for example, it is used to model the interaction of hadrons in short range gauge theories where coupling is mediated by the exchange of a massive scalar meson [1,2]. In atomic and molecular physics, it represents a screened Coulomb potential due to the cloud of electronic charges around the nucleus, which could be treated in the Thomas-Fermi approximation that leads to [3]

$$V(r) = -\frac{A}{r} e^{-\mu r},  \qquad (1)$$

where $\mu$ is the screening parameter and $A$ is the potential strength. This potential also describes the shielding effect of ions embedded in plasmas where it is called the Debye-Huckel potential [4]. It has also been used to describe the interaction between charged particles in plasmas, solids and colloidal suspensions [5]. The number of bound states of the Yukawa potential is always finite. However, due to the delicate nature of the resonances in the Yukawa potential, this subject did not receive adequate attention in the literature [6]. The solution of the Schrödinger equation for this potential has been investigated extensively in the past using various numerical and perturbative approaches since exact analytical solutions are not possible [7]. Despite the short-range behavior of the potential due to the decaying exponential factor $e^{-\mu r}$, the $r^{-1}$ singularity at the origin makes the task of obtaining accurate or even meaningful solutions a non-trivial and sometimes formidable task. Most of the perturbative and variational calculation found in the literature suffer from the limited accuracy when considering a wider range of potential parameters. Our approach constitutes a significant contribution in this regard. It is inspired by the J-matrix method [8] that handles this particular singularity not just accurately but in fact, exactly leaving the remaining non-singular and finite part to be easily treated numerically to the desired accuracy.

---
[*] Corresponding Author, email: haidari@mailaps.org

The J-matrix method is an algebraic method for extracting resonance and bound states information using computational tools devised entirely in square integrable bases. The total Hamiltonian is a sum two parts: a reference Hamiltonian $H_0$ and the remaining terms which are combined into an effective potential $U(r)$. The reference Hamiltonian is treated analytically and, thus, its contribution will be accounted for in full. As such, it must belong to the class of exactly solvable problems that could include singular interactions like $r^{-1}$ (e.g., the Coulomb potential). However, the effective potential $U(r)$ will be treated numerically. Therefore, for meaningful results, it must be non-singular, bounded everywhere and, preferably but not necessarily, short-range [9]. Now, the discrete $L^2$ bases used in the calculation and analysis are required to carry a tridiagonal matrix representation for the reference wave operator. The use of discrete basis sets offers considerable advantage in the calculation because it is an algebraic scheme that requires only standard matrix technique. The real power of our approach comes from the unique feature of J-matrix method that allowed us to isolate the $r^{-1}$ singularity of the Yukawa potential and absorb it into the reference Hamiltonian where it is treated exactly analytically [10]. Moreover, the choice of basis, which supports a tridiagonal matrix representation for the reference wave operator, results in a very efficient, stable and highly accurate means for evaluating the matrix elements of the remaining regular part (the effective potential) using Gauss quadrature integral approximation [11]. Working in a finite but highly accurate matrix representation of the total Hamiltonian, we study the bound states spectrum and resonance energies using the method of complex scaling [12]. We investigate the behavior of these energy eigenvalues and follow their trajectories in the complex energy plane as we vary the two Yukawa potential parameters. We arrive at the remarkable observation that as the potential parameters vary, bound states move up the energy spectrum until they reach the continuum where they experience transition into scattering states [13]. This phenomenon is demonstrated graphically and shown as video animation. To illustrate the utility and accuracy of our approach, bound states and resonances energies are compared satisfactorily with those obtained by other studies. Additionally, we present results for a range of values of the potential parameters that were never probed before. In the following, we start developing the tools of the approach and then apply them with the help of the J-matrix and complex scaling methods to arrive at our findings.

The time-independent Schrödinger equation for a particle of mass $m$ and charge $q$ in the combined field generated by the Coulomb potential and a spherically symmetric potential $V(r)$ reads as follows

$$(H-E)|\psi\rangle = \left[ -\frac{1}{2}\frac{d^2}{dr^2} + \frac{\ell(\ell+1)}{2r^2} + \frac{Z}{r} + V(r) - E \right]|\psi\rangle = 0, \quad (2)$$

where we have used the atomic units $\hbar = m = q = 1$ and length is measured in units of $a_0 = 4\pi\epsilon_0\hbar^2/mq^2$. In our case, $V(r)$ is the Yukawa potential given by Eq. (1). The combination $\frac{Z}{r} + V(r)$ in Eq. (2) is a Helmann-type potential. Because of the $r^{-1}$-type singularity of this potential, which is easily handled by the J-matrix method, we absorb it in the reference Hamiltonian by writing it as follows

$$H_0 = -\frac{1}{2}\frac{d^2}{dr^2} + \frac{\ell(\ell+1)}{2r^2} + \frac{\mathcal{Z}}{r}, \quad (3)$$

where $\mathcal{Z} = Z - A$. Therefore, the effective potential $U = H - H_0$, which will be treated numerically, has the following form

$$U(r) = -\frac{A}{r}\left(e^{-\mu r} - 1\right). \tag{4}$$

One can easily verify that this effective potential is regular and bounded everywhere. Consequently, we can evaluate its contribution (matrix elements) in a suitable $L^2$ basis to the desired accuracy using any preferred numerical integration scheme. The $r^{-1}$-type singularity in the reference Hamiltonian and the requirement that its matrix representation be tridiagonal dictate that the J-matrix basis used should be the "Laguerre basis" defined as [10]

$$\phi_n(x) = a_n x^\alpha e^{-x/2} L_n^\nu(x); \qquad n = 0,1,2,.. \tag{5}$$

where $x = \lambda r$, $\lambda > 0$, $\alpha > 0$, $\nu > -1$, $L_n^\nu(x)$ is the Laguerre polynomial, and $a_n$ is the normalization constant $\sqrt{\lambda \Gamma(n+1)/\Gamma(n+\nu+1)}$. Choosing $\alpha = \ell+1$ and $\nu = 2\ell+1$ gives the following tridiagonal matrix representation for $H_0$ [14]

$$\frac{8}{\lambda^2}(H_0)_{nm} = \left(2n+\nu+1+\frac{8\mathcal{Z}}{\lambda}\right)\delta_{n,m} + \sqrt{n(n+\nu)}\delta_{n,m+1} + \sqrt{(n+1)(n+\nu+1)}\delta_{n,m-1}. \tag{6}$$

In the manipulation, we used the differential equation, differential formula, three-term recursion relation, and orthogonality formula of the Laguerre polynomials [15]. Now, the only remaining quantity that is needed to perform the calculation is the matrix elements of the effective potential $U(r)$. This is obtained by evaluating the integral

$$U_{nm} = \int_0^\infty \phi_n(\lambda r) U(r) \phi_m(\lambda r) dr = \lambda^{-1} a_n a_m \int_0^\infty x^\nu e^{-x} L_n^\nu(x) L_m^\nu(x) \left[x U(x/\lambda)\right] dx. \tag{7}$$

The evaluation of such an integral for a general effective potential is almost always done numerically. We use the Gauss quadrature approximation [11], which gives

$$U_{nm} \cong \sum_{k=0}^{N-1} \Lambda_{nk} \Lambda_{mk} \left[\varepsilon_k U(\varepsilon_k/\lambda)\right], \tag{8}$$

for adequately large integer $N$. $\varepsilon_k$ and $\{\Lambda_{nk}\}_{n=0}^{N-1}$ are the $N$ eigenvalues and corresponding eigenvectors of the $N \times N$ tridiagonal basis overlap matrix $\langle \phi_n | \phi_m \rangle$. Therefore, the reference Hamiltonian $H_0$ in this representation, which is given by Eq. (6), is accounted for in full. On the other hand, the effective potential $U$ is approximated by its matrix elements in a subset of the basis. In this letter, we limit our investigation to the structure and dynamics of bound states and resonances and are contented with a finite dimensional representation of the total Hamiltonian. Moreover, we consider only pure Yukawa coupling (i.e., Coulomb-free interaction with $Z = 0$).

Direct study of bound states and resonances is usually performed in the complex energy plane ($E$-plane). In such studies, these states are identified with the poles of the Green's function, which is defined formally in the $E$-plane as $G(E) = (H - E)^{-1}$. For systems with self-adjoint Hamiltonian, resonances are located in the lower half of the second sheet of the complex energy plane. One of the methods of investigation of resonances in the $E$-plane is the complex scaling (a.k.a. complex rotation) method [12]. In this method, the radial coordinate gets transformed as $r \to re^{i\theta}$, where $\theta$ is a real angular parameter. The effect of this transformation on the pole structure of the Green's function consists of the following: (1) the discrete bound state spectrum that lies on the negative energy axis remains unchanged; (2) the branch cut (the discontinuity) along the real positive energy axis rotates clockwise by the angle $2\theta$; (3) resonances in the lower half of the complex energy plane located in the sector bound by the new rotated cut line and the positive energy axis get exposed and become isolated. However, due to the finite size of the basis used in the calculation, the rotated cut line gets replaced by a string of interleaved poles and zeros of the finite Green's function, which tries to mimic the cut structure. One can easily show that the complex scaling transformation $r \to re^{i\theta}$ of the total Hamiltonian in configuration space is equivalent to the transformation of its matrix elements by changing the scale parameter as $\lambda \to \lambda e^{-i\theta}$.

The first video [16] that accompanies our work is an animated plot (a series of accurate graphical representations) in the $E$-plane of the calculated energy eigenvalues of the finite $N\times N$ total Hamiltonian matrix without complex scaling ($\theta = 0$). In the animation, we took $\ell =1$, $\mu = 5$ (a.u.) and $A$ was varied from 80 to 0 (a.u.). The string of points on the positive real line is the set of energy eigenvalues corresponding to the discretized continuum line. In the beginning where $A = 80$ there is only one bound state in view with an energy near $E = -3.2$ (a.u.). As the bottom of the effective potential well rises (with $A$ decreasing) this energy eigenvalue starts shifting up in the spectrum by moving to the right until it gets "absorbed" into the continuum. Decreasing $A$ further, a new bound state comes into view from left moving to the right until it eventually becomes embedded into the continuum as well. This process continues. In the second animation [17], we show the remarkable fact that bound states do not just get absorbed into the continuum but, in fact, experience a transition into scattering states [13]. To illustrate this, we use the complex scaling method described briefly above. It is obvious from the second video that the bound states energy eigenvalues that move to the right half of the $E$-plane do not rotate with the continuum line but follow their own resonance trajectories. These trajectories are stable against variations in all computational parameters (the rotation angle $\theta$, basis scaling parameter $\lambda$, etc.). Figure 1 is a plot of these trajectories for the three states contributing to the animation. We have also seen the same phenomenon repeated when fixing the strength of the potential $A$ while varying its range parameter $\mu$. Video animations and figures for this case are available upon request from the corresponding author.

To illustrate further the utility and accuracy of our approach, we use it to calculate bound states and resonance energies for a given set of physical parameters where we can compare our results with those obtained elsewhere [18-20]. Our calculation strategy is as follows. For a given choice of physical parameters, we investigate the stability of calculated eigenvalues that correspond to bound states and/or resonances as we vary the scaling parameter $\lambda$ until we reach a plateau in $\lambda$ [21]. Then to improve on the accuracy of the results, we select a value of $\lambda$ from within the plateau and increase the dimension of the space $N$ until the desired accuracy is reached. Table 1 lists some of the bound states and resonance energies where our results are compared with available numerical data [18-20] satisfactorily. However, Table 2 contains results that are unique to our approach since the values of the potential parameters and range of energies fall outside the applicability of most perturbative and variational calculations found elsewhere. More results could be found in [22]. Finally, this approach could easily be generalized to handle other short-range potentials with $r^{-1}$ singularity. For example, we are currently involved in extending it to the Hulthén potential.

ACKNOWLEDGMENTS

The authors acknowledge the support provided by the Physics department at King Fahd University of Petroleum & Minerals.

**TABLE CAPTIONS:**

**Table 1:** Bound states and resonance energies for the Yukawa potential with the given parameters. The strength of the potential is normalized as $A = 1.0$. Our results are compared with those of others [18-20]. All values are in atomic units.

**Table 2:** Bound states and resonance energies for a wider range of parameters of the Yukawa potential. All values are in atomic units.

**Table 1**

| [Ref.] | $\ell$ | $\mu$ | $E$ [Ref.] | $E$ [this work] |
|---|---|---|---|---|
| [21] | 0 | 0.10 | −0.407058030613403156754507070 | −0.407058030613 |
|  |  | 0.10 | −0.049928271331918889234996681 | −0.049928271332 |
|  |  | 0.10 | −3.20804674469025 E-3 | −3.208046744693 E-3 |
|  | 1 | 0.20000 | −4.1016465307840 E-3 | −4.101646530802 E-3 |
|  | 2 | 0.08 | −3.248360428751 9935 E-3 | −3.248360428763 E-3 |
| [22] | 1 | 0.112 | −5.003810 E-5 | −5.008834873206 E-5 |
|  |  | 0.113 | +1.631328 E-5 −i 1.0250 E-6 | +1.690357641732 E-5 −i 1.862588128150 E-6 |
|  |  | 0.122 | +4.21637295 E-4 −i 3.2519395 E-4 | +4.214837407095 E-4 −i 3.251672270506 E-4 |
| [23] | 1 | 0.11260485 | −3.9326971 E-2 | −3.932698311615 E-2 |
|  |  | 0.11324071 | −3.8977839 E-2 | −3.897785063740 E-2 |
|  |  | 0.11324071 | +3.0964751 E-05 −i 4.6283598 E-06 | +3.096253670542 E-5 −i 4.640977622365 E-6 |
|  |  | 0.22007434 | −1.8679864 E-5 | −1.867969736495 E-5 |
|  |  | 0.22131705 | +1.3758648 E-4 −i 1.5287316 E-05 | +1.375843424574 E-4 −i 1.528916191194 E-5 |
|  | 2 | 0.091 | −7.767555 E-5 | −7.767498169649 E-5 |
|  |  | 0.092 | +1.410647 E-4 −i 1.1700 E-6 | +1.410607617484 E-4 −i 1.170284982249 E-6 |
|  |  | 0.20719391 | −7.4344785 E-3 | −7.434478537514 E-3 |
|  |  | 0.20719391 | +7.5614845 E-4 −i 5.3974260 E-4 | +7.561484464990 E-4 −i 5.397425935464 E-4 |
|  |  | 0.27206629 | −2.8843407 E-5 | −2.884314042012 E-5 |
|  |  | 0.27523818 | +8.7320680 E-5 −i 3.5919478 E-7 | +8.732080984385 E-5 −i 3.591924399130 E-7 |
|  | 5 | 0.32431943 | +2.8791297 E-5 −i 4.9081260 E-12 | +2.879132724089 E-5 −i 4.651948920811 E-12 |
|  |  | 0.39266564 | +9.4978169 E-4 −i 2.4497104 E-4 | +9.497816886083 E-4 −i 2.449710308446 E-4 |
|  | 10 | 0.34283428 | −5.5563066 E-06 | −5.556304541175 E-6 |
|  |  | 0.34495566 | +7.2188283 E-6 −i 0.000 | +7.218866481722 E-6 −i 1.786030250706 E-15 |

**Table 2**

| | $\ell$ | $\mu$ | $A$ | $E$ |
|---|---|---|---|---|
| Bound states | 0 | 1.0 | 30 | −420.7340520442 |
| | | | | −85.29472078885 |
| | | | | −25.84943144298 |
| | | | | −7.656289423403 |
| | | | | −1.517703985470 |
| | | | | −0.020763186246 |
| | 1 | 0.5 | 20 | −40.59698542831 |
| | | | | −13.63156417158 |
| | | | | −4.917728174528 |
| | | | | −1.546320478514 |
| | | | | −0.285657326503 |
| | 2 | 0.2 | 10 | −3.751512770069 |
| | | | | −1.494005746764 |
| | | | | −0.566900519026 |
| | | | | −0.168517340167 |
| | | | | −0.019233342003 |
| Resonances | 2 | 10 | 108 | 3.06069726414 −i 0.06221472755 |
| | | | 170 | 2.48495743898 −i 0.08049696875 |
| | | | 248 | 1.5027589904 −i 0.0376796110 |
| | 5 | 2 | 110 | 2.4208899272 −i 0.0768589832 |
| | | | | 3.0139488985 −i 3.6049527387 |
| | | | 170 | 1.024443520 −i 0.003598625 |
| | | | | 2.236370014 −i 2.381751848 |
| | | | 200 | 1.359477152 −i 0.029358570 |
| | | | | 1.815292940 −i 2.549254736 |
| | 10 | 1 | 200 | 1.057885231 −i 0.000001090 |
| | | | | 1.651827352 −i 4.794059401 |
| | | | | 3.065622630 −i 2.731016510 |
| | | | | 3.264518294 −i 0.639296036 |

**FIGURE CAPTION:**

**Fig. 1:** A plot in the complex energy plane of the trajectories of the energy eigenvalues (in atomic units) corresponding to the three states (bound and resonance) contributing to the animation. The physical parameters (in atomic units) are $\ell = 1$, $\mu = 5$ and $A$ was varied from 80 to 0.

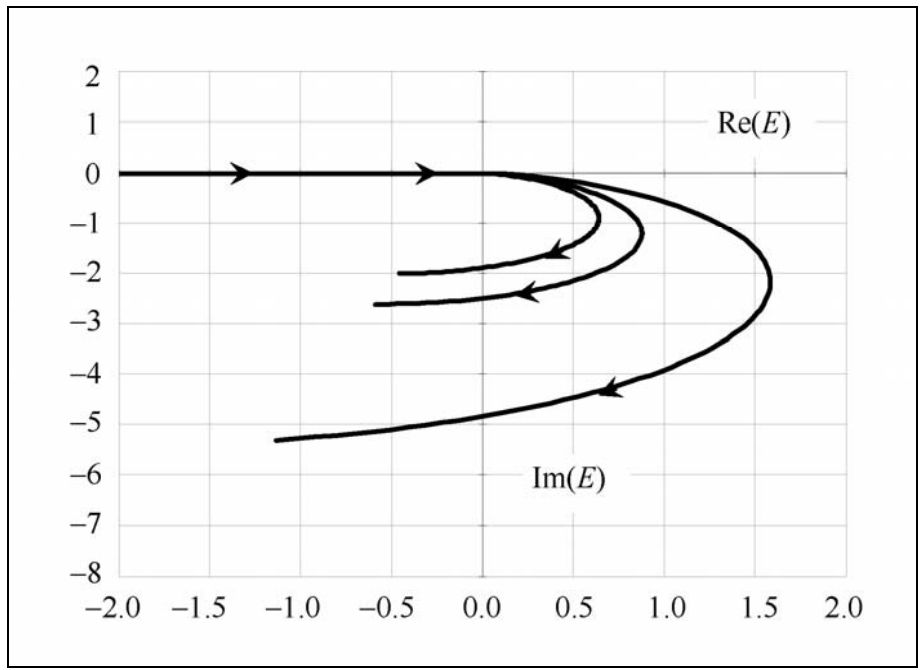

**Fig. 1**